# A Fourier ($k$-) space design approach for controllable photonic band and localization states in aperiodic lattices

**Subhasish Chakraborty [a*], Michael C. Parker [b], Robert J. Mears [c]**

[a] Microelectronics Research Centre, Cavendish Laboratory, Department of Physics, University of Cambridge, Madingley Road, Cambridge CB3 0HE, UK

[b] Fujitsu Laboratories of Europe Ltd., Columba House, Adastral Park, Ipswich IP5 3RE, UK

[c] Pembroke College, Trumpington Street, Cambridge CB2 1RF, UK

[*]Corresponding author, Present address: School of Electronic & Electrical Engineering, University of Leeds, Leeds LS2 9JT, UK, Tel: +44 (0)113 343 2017, Fax: +44 (0)113 343 7265, E-mail: S.Chakraborty.98@cantab.net

**Abstract**

In this paper we present a systematic study of photonic bandgap engineering using aperiodic lattices (AL's). Up to now AL's have tended to be defined by specific formulae (e.g. Fibonacci, Cantor), and theories have neglected other useful AL's along with the vast majority of non-useful (random) AL's. Here, we present a practical and efficient Fourier space-based general theory to identify all those AL's having useful band properties, which are characterized by well-defined Fourier (i.e. lattice momentum) components. Direct control of field localization comes via control of the Parseval strength competition between the different Fourier components characterizing a lattice. Real-space optimization of AL's tends to be computationally demanding. However, via our Fourier space-based simulated annealing inverse optimization algorithm, we efficiently tailor the relative strength of the AL Fourier components for precise control of photonic band and localization properties.

## 1. Introduction

In the design of photonic crystals used in photonic integrated circuits and dense wavelength division multiplexing (DWDM) systems, increasing importance is being placed on careful control of the transmission characteristics, which derive from the device photonic band properties. One particular requirement is high transmission and high resolution operation. There is thus a high premium on the ability to engineer the band properties of a photonic crystal in a systematic manner. Photonic crystals [1] are usually characterized by a well-defined lattice periodicity. In the simplest case, for a periodic 1D lattice, where the wavelength of the electromagnetic (EM) radiation is twice the optical lattice constant ($\Lambda$), Bragg-resonance occurs and a bandgap forms in the dispersion characteristic, which results in a dip in the spectral transmission characteristic, but no useful photon localization. In order to achieve field localization or fine tuning of the band properties, photonic bandgap (PBG) engineering [2,3] is needed, which requires breaking of the lattice periodicity thorough the introduction of single or multiple defects (missing or extra scattering sites) [4-7]. In real-space, they determine the localized cavities storing energy at those defect sites. In general, when an EM field is localized in a lattice-based cavity, the associated cavity mode provides the basis for useful filtering functionality. In finite lattices, such modes couple to propagating modes to appear as a dramatically enhanced transmission in comparison with other frequencies in the vicinity [4]. Although Bragg resonances are responsible for determining the photon propagation through a lattice, it is worth mentioning that fundamentally, such localization phenomena of the EM wave can also be related to Bragg resonances.

Conventional PBG engineering has tended to be based mostly on a ‘forward’ process of defining a lattice structure and then determining the band properties, e.g., through the use of constraining formulae such as the Fibonacci or Cantor-based quasi-periodic lattices [8-10]; through coupled-cavity structures, proposed by Yariv *et al*. [6]; or by intuition (accumulated design experience) and trial-and-error, such as the high-Q cavity structures reported by Akahane *et al*. [11]. There has also been a growing interest in the 'inverse' process of determining appropriate aperiodic photonic lattices from the desired scattering properties (functionality) [12-16]. With the exception of quasiperiodic lattices, which can form stop-bands considerably away from the conventional Bragg frequency, these techniques modify the spectral functionality mostly within or around the Bragg stop-band. Also, in most cases (both 'forward' and 'inverse') the underlying principle is the direct optimization of the band properties by characterization of the lattices in real-space. Even for a moderate number of scattering sites and defects, these methods become computationally extremely demanding and complex to generalise beyond 1D. While PBG engineering renders the lattice essentially aperiodic [17,18] in the most general sense, it is highly unlikely that a randomly chosen aperiodic lattice (AL) will have any useful field localization or band properties. Hence, the question is how to identify the relatively small number of useful AL's (in terms of their band properties) from the very large number of possible AL's?

In this paper we present a general theory and systematic design tool to reveal a previously unknown landscape of essentially aperiodic lattices, which show useful and novel spectral functionalities, e.g., tuneable bandgaps and multiple localized states. Previous theories have discarded these useful AL's along with the vast majority of non-useful (random) AL's, because they have characterized photonic lattices only with regular periodicity of the real space lattice. However, as we explain in our paper, we have adopted a Fourier-space (*k*-space) approach [19], where AL's having useful band properties are characterized by well-defined spatial frequency (i.e., lattice momentum) components. We have used a discrete Fourier transform (DFT)-based inverse optimization algorithm to tailor the Fourier components of an AL to match that of a "target" function; the target function, also, of course, being defined in Fourier-space and equating to the desired band properties. Overall, the entire scheme turns out to be more practical and efficient in achieving desired and controllable photonic band functionality. Indeed, we demonstrate that AL's [20-23] inherently offer the most flexible platform to achieve novel spectral functionalities. In this context, our method provides answers to questions such as: For a given overall length and lattice constant $\Lambda$ of a 1D lattice, what are the various defect combinations that produce discontinuities along the angular frequency *w*-axis corresponding to various discrete wavevector points between, say, $k_{BR}= \pi/\Lambda$ (edge of the Brillouin zone) and $0.5k_{BR}$? Or more importantly in the context of optical telecommunications, e.g., DWDM components: What defect combinations for this lattice will provide localized photon states (i.e., high photon transmission) at desired optical frequencies, useful for multi-wavelength narrowband optical filters and tuneable semiconductor lasers? Although some of the resulting aperiodic structures may appear 'intuitive' and 'expected' from the optimisation algorithm, we emphasise the fact that the scattering positions in a lattice structure are subtly balanced to provide both the requisite photonic bandgap and also the localised states within that bandgap. This is because the presence of a bandgap requires the appropriate constructive phase relationships between the Fourier components of the lattice, while the introduction of defects must be controlled such that bandgap impairment is minimised, while localised states appear both within that bandgap, but also with reasonable finesse. Hence, the resulting structures should not be considered as post-facto 'obvious' or 'trivial' results, but as delicately engineered optimised compromises.

This paper is organized as follows: In section 2 we establish a powerful link between PBG engineering and digital signal processing (DSP) to enable a Fourier transform (FT) based inverse optimization process, e.g., for photon localization. In section 3 we present the mathematical framework for our inverse optimization technique using simulated annealing (SA). Section 4 provides results and discussions for shifting the photonic bandgap away from the Bragg frequency, and for photon localization. Finally, in section 5 we present some conclusions.

## 2. Photonic bandgap, EM wave localization and digital signal processing

The underlying physics behind the PBG effect and EM wave localization is Bragg resonance between the wavevector $k$ of EM radiation and lattice momentum $G$. In the simplest case of a periodic 1D lattice, composed of layers of two different materials with geometric thicknesses, $H$ and $L$, and refractive indices, $n_H$ and $n_L$, respectively, when Bragg's law $\vec{k}_{\mathrm{BR}} . \hat{G} = G_{\mathrm{BR}} / 2 = p / \Lambda$ is satisfied for a photon of wavevector $\vec{k}_{\mathrm{BR}}$, (with $\hat{G}$ the unit lattice momentum vector), a discontinuity (Bragg stop-band) forms. This equates to a dip in transmission through the structure about an angular frequency $w_{\mathrm{BR}}$ (the Bragg frequency) given by $w_{\mathrm{BR}} = ck_{\mathrm{BR}}$, where $c$ is the vacuum speed of light. We note the existence of the dominant Fourier component at an optical spatial frequency (momentum) given by $G_{\mathrm{BR}} = 2p / \Lambda$, where $\Lambda = n_H H + n_L L$ is the periodicity in optical space. In our paper, real-space and reciprocal-space variables are all assumed the appropriate optical quantity [24] (i.e., optical space or path-length is given by $x = \int n(l) dl$ where $l$ is geometric real-space, and optical reciprocal-space is given by $G = 2p / x$), unless otherwise stated. It is widely known in DSP that the DFT of a real series produces a symmetric amplitude spectrum about the highest sampling frequency (known as the Nyquist frequency). This principle, taken together with the symmetry observed in the spectral responses of 1D quasiperiodic superlattices [9], suggests that the scattering sites of any real-space AL can be regarded as samples with a spatial frequency less than the highest spatial frequency $G_{\mathrm{BR}}$ (the Nyquist frequency). Hence, the FT of an AL defined by a set of scattering sites $\{x_p\}$ will give a set of spatial frequency components $\{G_q\}$ that forms a symmetric spectrum about the highest spatial frequency $G_{\mathrm{BR}}$, where the lattice reciprocal-space variable $G$ is the Fourier conjugate to the real-space variable $x$. Generalising the Bragg resonance condition between all of these spatial frequency components $\{G_q\}$ and the set of EM angular frequencies $\{w_q = cG_q / 2\}$ gives us a qualitative determination of the spectral transmission (reflection) characteristic for EM wave propagation through the AL, indicated in Fig. 1. The strength of the Bragg resonance at that frequency is controlled by the Fourier coefficient amplitude $\bar{e}\{G_q\}$ given by the FT of the scattering dielectric function $e\{x_p\}$. Clearly, this procedure results in transmission dips centred on the set of frequencies $\{w_q\}$, with symmetry about $w_{\mathrm{BR}} = cG_{\mathrm{BR}} / 2$. We note that the localization of the EM wave is determined by those wavevectors of the EM wave spectrum, which have their Bragg resonating partner in the Fourier spectrum of the scattering dielectric function. Localization of the EM wave inside a lattice can be thought of as originating from the interaction of different Bragg resonances, i.e., those finite Fourier components characterising the lattice. Multiple Bragg resonances can therefore be used as the basis of controlling EM localization in a fashion not available from a conventional periodic lattice (which exhibits only a single Bragg resonance). The FT-basis of the localization phenomenon, as qualitatively outlined above, allows us the additional benefit of exploiting DSP techniques, as explained above. A DFT of a (purely-real) distribution produces a symmetric Fourier spectrum about $G_{\mathrm{BR}}$, where each component is responsible for a Bragg resonance. Due to considerations of power conservation (i.e., Parseval's theorem), the wave localization properties of the lattice are determined by the strength competition between those Fourier components. Overall, this is an initial clue that the transmission characteristic of a photonic lattice is closely related to its spectral-distribution characteristics, i.e., the FT of its real-space structure (equivalent to the first Born approximation [25]). This FT link is the basis of our inverse design approach.

### 3. Inverse optimization

For mathematical convenience, the following DFT-based inverse analysis has been conducted in 1D only. However, we note that since the analysis is FT-based, it can be extended to higher-dimension photonic lattices. We consider a binary relative permittivity lattice structure $e\{x_p\} = e_{\text{ave}} + \Delta e\{f_p\}$ of $N$ sites located at the set of positions $\{x_p\}$ in optical space. Each site is taken to be either $n_H$ or $n_L$, of respective geometric thicknesses, $H$ and $L$. We use $e_{\text{ave}} = (n_H^2 + n_L^2)/2 = n_{\text{ave}}^2$ and $\Delta e = (n_H^2 - n_L^2)/2$. The set of parameters $\{f_p\} = \pm 1$ represents the binary lattice function (analogous to the Bravais lattice function for a periodic structure), where $f_p$=+1 equates to a high refractive index $n_H$, and $f_p = -1$ to the low refractive index $n_L$, so that defects inside the lattice can be controlled simply by manipulating the polarity $f_p$ of any individual site. Consider an arbitrary $h^{\text{th}}$ configuration of the photonic lattice denoted by $e_h\{x_{\text{p}}\}$. Taking the DFT of this $h^{\text{th}}$ configuration, yields its set of discrete spectral components $\bar{e}_h\{G_q\}$ given by:

$$\bar{e}_h\{G_q\} = \frac{\Delta e}{N} \sum_{p=1}^{N} f_p e^{-iG_q x_p} \ . \qquad (1)$$

Periodic boundary conditions define the set of discrete spatial frequencies $G_q$, given by

$$G_q = \frac{2q}{N} G_{\mathrm{BR}}, \quad q = 1,2,\ldots.. N/2, \tag{2}$$

where $G_{\mathrm{BR}}$ denotes the highest (Nyquist) spatial frequency $2p/\Lambda$. The maximum value of $q=N/2$ for the following analysis is sufficient due to the symmetric redundancy in the spectrum about the Nyquist frequency, discussed above. We can tailor the Fourier components distribution, given by Eqs. (1) and (2) according to our need, and a subsequent inverse FT will generate a real-space lattice configuration. Various band properties, e.g., the strength of bandgaps or localized states, depend only on the amplitude of the associated Fourier component, with the phase characteristic, therefore, being a degree of freedom (i.e., arbitrary). This makes the inverse calculation analogous to the calculation of a computer-generated hologram (CGH) [26], with the formidable computational challenge of identifying those purely real CGH solutions from within the factorially-large search space. We use a non-deterministic simulated annealing (SA) optimization algorithm to search this configuration space. By controlling the defects in the lattice its configuration is optimized when the cost function $E$ is minimized, where

$$E(e_h) = \sum_{q=1}^{N/2} \left[ \left| \bar{e}_{\mathrm{target}} \{G_q\} \right| - \left| \bar{e}_h \{G_q\} \right| \right]^2, \tag{3}$$

describes the ‘error’ between the lattice spectral response and the target spectral distribution $\bar{e}_{\mathrm{target}}\{G_q\}$. We note that multiple runs of the SA algorithm will tend to find different solutions, each is close to an overall global optimum in cost-space. From a practical point of view, the functionality of these solutions tends to be indistinguishable. As part of the design process, the appropriate value for $N$ is adopted to reflect the fabrication resolution and the overall physical size of the lattice, with $G$ existing in a quasi-continuous space as $N$ tends to infinity. Having generated an AL, we check that its transmission properties (i.e., band properties) are in agreement with the desired transmission characteristics. Conventional methods for this forward determination of the transmission characteristics of an AL use either transfer matrix (TM) methods, finite-difference time-domain (FDTD) approaches, or eigen-mode expansion (EME) techniques. However, we use the Fourier transform Eq. (4), which has been derived in Ref. [24] by solving the Ricatti equation for the scattering coefficient between a pair of forward- and backward-propagating coupled-modes, for quick and efficient (yet reasonably accurate) calculation of the transmission characteristics of our AL's.

$$t(k) = \operatorname{sech}\left[\left|\frac{1}{4n_{\text{ave}}^2}\int_{-\infty}^{\infty}\left(\frac{\partial e(x)}{\partial x}\right)e^{-2ikDx}\,dx\right|\right]. \tag{4}$$

$D$ is a modified Debye-Waller factor required to avoid the well-known phase accumulation error. The $k$-space design was carried out using a software program written in MATLAB$^{\text{TM}}$, taking advantage of the efficient FFT algorithm, so that the full SA optimisation for a particular AL with $N$=112 took less than 1 minute using a Pentium IV processor with 2.8GHz clock frequency and 512MB RAM. For larger $N$, the number of operations for the FFT scales as $N\log N$, and the overall SA optimisation scales accordingly. Higher dimensionality $d$ AL's simply scale as $dN^d \log N$.

## 4. Results and discussions

Having designed an AL, a commercial software package (FIMMPROP-3D) [27] was used to verify the performance of the resulting AL. The aperiodic binary photonic bandgap lattice was assumed to be realised in a rib-type waveguide by implementing the required changes in relative permittivity through binary modulation of the waveguide width. The effective refractive indices of the fundamental mode in the two widths of the waveguide were found to be $n_{\mathrm{H}}$=2.6 and $n_{\mathrm{L}}$=2.42, respectively. The optical path length for each waveguide-section was chosen to be a quarter Bragg-wavelength, with the Bragg wavelength $l_{\mathrm{BR}}$ =1550nm, so that the two geometric lengths are $H=l_{\mathrm{BR}}/4n_{\mathrm{H}}$ ≈149nm and $L=l_{\mathrm{BR}}/4n_{\mathrm{L}}$ ≈160nm. This gives the smallest geometric spatial period in the lattice as $\Lambda' = H + L = 309$nm (i.e., $\Lambda = n_{\mathrm{H}}H + n_{\mathrm{L}}L = 775$nm in optical space). Employing $N$=112 unit-cells, with the normalised DFT Nyquist frequency ($G_{\mathrm{BR}}$) corresponding to $N/2 = 56,$ we have 28 discrete points (indexed by the DFT variable $q$) between the edge of the Brillouin zone (corresponding to $k_{\mathrm{BR}}$, $q$=56) and 0.5$k_{\mathrm{BR}}$, where $q$=28. To illustrate this, we present results for the first five points corresponding to $q$=56,55,54,53, and 52, respectively. We emphasize, though, that the lattice can be designed to exhibit a stop band ‘on demand’ at any position corresponding to $q$=1 through to $q$=56; however, space constraints allow us to show in Fig. 2 (i-v) only examples for the first five positions (the corresponding lattice configurations are presented in Table I). Figure 2(i) shows the spectrum for a conventional periodic lattice with a bandgap at the edge of the Brillouin zone (i.e., $q$=56). Transmission characteristics in Fig. 2 (ii-v) show the symmetric photonic stop gaps being shifted in incremental steps of about 0.02$f_{\mathrm{BR}}$, (i.e., $2f_{\mathrm{BR}}/N$) away from the Brillouin zone edge, by using defects in well-defined locations in the photonic lattice. We also note the occurrence of resonant peaks between the two symmetric photonic stop bands. These resonant peaks form as a result of interference of different Bragg stop-bands, i.e., those Fourier components which make the lattice. Precise control of these peaks, important for DWDM components (e.g., when realised within the cavity of a semiconductor laser to enable tuning [28]) thus requires tailoring of the relative strength of the different Bragg stop-bands. An important example is the so-called single-defect (0.5Λ) Fabry-Perot-type lattice [29] (used in a quarter-wave shifted DFB laser), corresponding to the $q$=55 AL. The high-frequency band-edge transmission peak of the lower bandgap (at $0.98f_{BR}$), and the low-frequency band-edge transmission peak of the upper bandgap (at $1.02f_{BR}$) overlap to form a single, very narrow resonant (localized) transmission peak at the Bragg frequency. In the example spectra of Fig. 2 (ii-v) we haven't tried controlling these peaks, as the intention is more to demonstrate the controlled tunability of Bragg stop-bands. By introducing ever more defects, Fig. 2 shows a progression from a periodic to ever-more aperiodic lattices, with the bandgap being shifted further away from the Brillouin zone edge. In so doing, the number of scatterers is reduced, so that the overall bandgap strength is reduced, but we note that we can straightforwardly increase the PBG bandgap strength again by simply increasing $N$, or the refractive index contrast $\Delta e$.

As a design example of an aperiodic lattice, with precisely defined wave localization properties, we have chosen to modify the spectral response, reported in Foresi *et al*. [4], from a single resonant high-Q defect state to a doubly resonant system with two high-Q defect states within a wide photonic stop band. In order to achieve this modification in the same system, we need to identify a real-space lattice with two properties; first, the lattice must have the same number (eight) of refractive index contrast elements (e.g., etched air holes in a Si-waveguide), i.e., $N$=16; second, the length of any defect is 0.5Λ. We also used silicon-silicon dioxide as our material system, which provides a sufficiently large contrast (the refractive index difference $\Delta n$ is about 2) so that the optical wave is strongly confined. The waveguide cross-section can, therefore, also be made very small (in this case 0.5 μm wide and 0.26 μm thick), which is useful in microphotonic integrated circuits at optical telecommunication wavelengths requiring microscale optical elements. The resulting device transmissivities in Fig. 3 were simulated using FIMMPROP-3D, discussed previously.

Figure 3 (i-b) shows the DFT spectrum for the single defect lattice shown in Fig. 3 (i-a), which produces a single high-Q resonant state within the photonic stop band, as seen in Fig. 3 (i-c). The Fourier amplitude is zero at the Nyquist frequency point $G_{BR}$, with the strongest Fourier components existing at the neighbouring Fourier positions, calculated as $0.875G_{BR}$ with its symmetric partner at $1.125G_{BR}$. The Bragg resonances, which originate from these main two Fourier components, interfere to form a very narrow (i.e., high-Q) transmission peak at the conventional Bragg frequency $f_{BR}$, i.e., a Fabry-Perot mode equivalent to Fig. 2 (ii). Figure 3 (ii) and (iii) show the DFT responses and the spectral transmission characteristics of two further lattices doped with multiple defects. The defects control the strength of the Fourier component at the Nyquist frequency, and, hence, the transmissivity at the Bragg-frequency. In accordance with Parseval's theorem, we note that the Fourier component at $G_{BR}$ in Fig. 3 (ii-b) and (iii-b) strengthens at the cost of the component amplitudes of the neighbouring Fourier locations. The strengths of the Fourier components (i.e., closely equivalent to reflection coefficients) determine the finesses associated with the two resonant peaks, and combined with cavity losses (e.g., lattice absorption, diffractive radiation) control the overall Q-values of the peaks. By inspection, the individual resonances in Fig. 3 (ii-c) and (iii-c) have lower finesses (and, hence, Q-values) than the single resonant peak of Fig. 3 (i-c). Figure 3 (iv-a) shows an AL forming two high-Q transmission peaks. With two defects of size $d$=0.5Λ, the two resonating defect states appear deep within the stop band at frequencies 0.92 $f_{BR}$ and 1.05 $f_{BR}$ (the asymmetry apparent with respect to the Bragg frequency being due to the additional lattice material dispersion as calculated by FIMMPROP-3D). Evident are the higher finesses of the peaks of Fig. 3 (iv-c) as compared with Fig. 3 (ii-c) and (iii-c). We also observe an interesting inverse relationship between the relative distance separating the spatial defects, and the frequency distance between the resulting resonating peaks. Such behaviour is perhaps to be expected, due to the reciprocal symmetries between the FT conjugate planes. The simulated field intensity patterns in the *xy*-plane corresponding to these modes and at the Bragg frequency have been plotted in Fig. 4. At the resonating frequencies each defect site acts as a cavity, where the electromagnetic wave is spatially localized. However, the evanescent tail of each mode can overlap and couple to the propagating modes resulting in an enhanced transmission for the corresponding resonant frequency. It also can be seen from Fig. 4., that although the spatial field distributions for both frequencies show maxima at the defect regions, there is a node between the defects for the low frequency resonance (Fig. 4 (a)). As might be expected, the symmetries exhibited by the spatial field distributions at the resonating frequencies have a close analogy with the bonding and antibonding wavefunctions well-known in solidstate physics. At the Bragg frequency, the high intensity of light entering from the left, but being reflected (i.e., not allowed to propagate) is clearly evident.

## 5. Conclusions

In conclusion, we have presented a Fourier-space based design approach for systematic control of the photonic band properties and the number of resonant states within a photonic stop band using aperiodic lattices. Using a simulated annealing optimisation algorithm, we have demonstrated that we can manipulate the Fourier components (constrained by Parseval's theorem) of an AL, to inverse design and engineer any desired bandgap and field localization characteristic. Our theory takes advantage of the analogies between PBG engineering and DSP, both underpinned with a common FFT formalism, to enable a computationally efficient inverse design methodology. Overall, in addition to the design of DWDM components, we believe this work will help PBG engineering find new applications, for example, with multiple defects introduced into a 3D photonic crystal fabricated using holographic lithography [30].

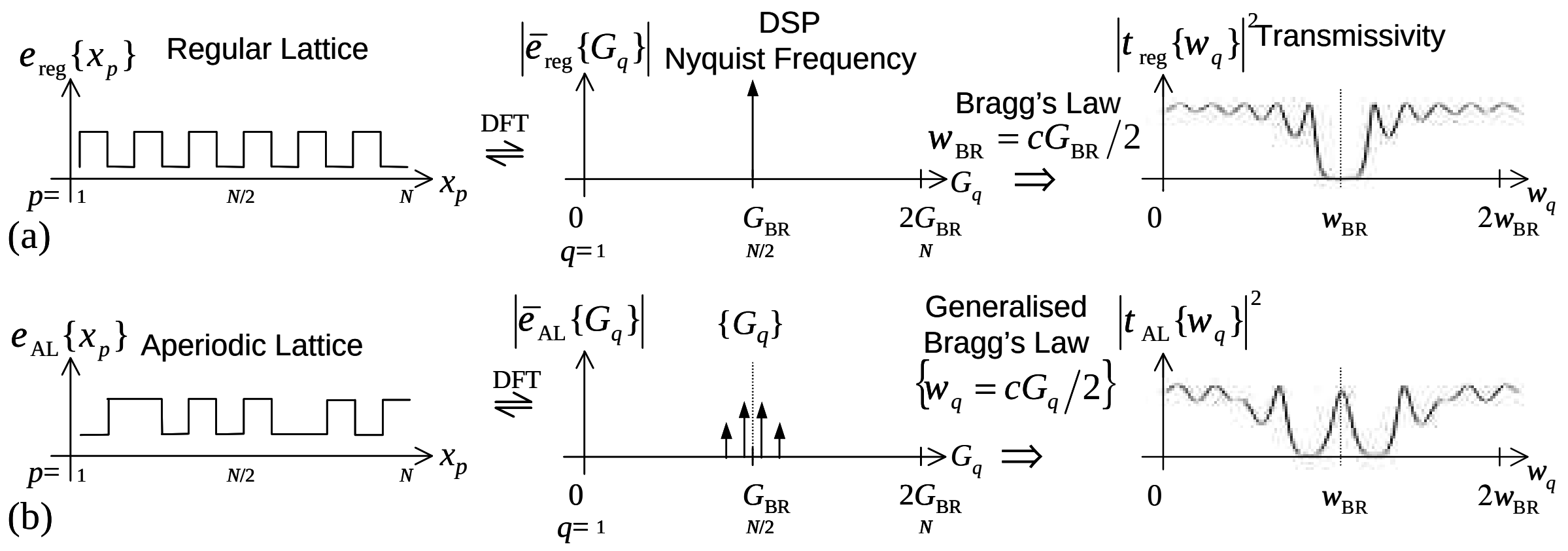

FIG. 1: Qualitative understanding of how we can combine simple rules of DSP with Bragg's Law, in order to predict the position of the dips in the electromagnetic transmission spectrum of any real AL.

Table 1. Optimised AL configurations

| Target Spatial Frequency index ($q$) | Detuning factor ($\Delta f/f_{BR}$) | High ($n_H$ )and Low ($n_L$,) RI combinations |
|---|---|---|
| 56 | 0 | $(HL)^{56}$ |
| 55 | 0.0179 | $(LH)^{28}$ $(HL)^{28}$ |
| 54 | 0.0357 | $(HL)^{13}$ $H^3$ $(HL)^{13}$ H $(HL)^{13}$ $H^3$ $(HL)^{13}$ L |
| 53 | 0.0535 | $(HL)^9$ H $(HL)^9$ L $(HL)^8$ $L^3$ $(HL)^8$ H $(HL)^9$ H |
| 52 | 0.0714 | $(HL)^6$ L $(HL)^6$ $H^3$ $L^2$ $(HL)^5$ $L^3$ $(HL)^6$ L $(HL)^7$ L $(HL)^6$ H $(HL)^6$ L $(HL)^7$ L |

Tuning of photonic stop bands in incremental steps away from the conventional Bragg frequency $f_{BR}$, using $N$=112 unit-cells, with the normalised DFT Nyquist frequency ($G_{BR}$) corresponding to $q$=$N/2 = 56$.

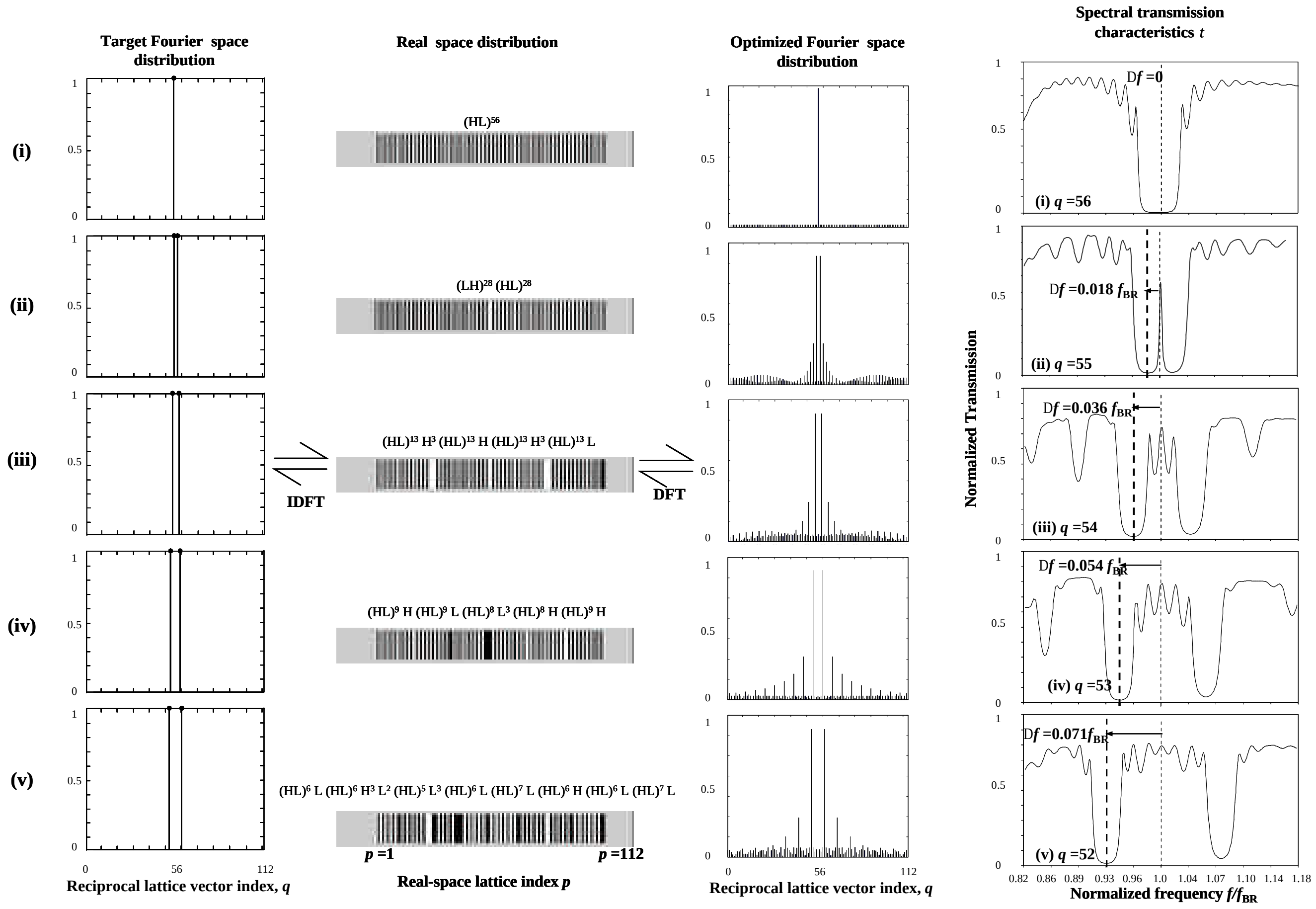

Figure 2. Optical transmission spectra (transmission $t$ versus normalized frequency $f/f_{BR}$), calculated using FIMMPROP-3D. Figures (i-v) show use of AL's to shift the PBG into the Brillouin Zone in steps of ~$0.02f_{BR}$ from the edge of the Bragg frequency $f_{BR}$.

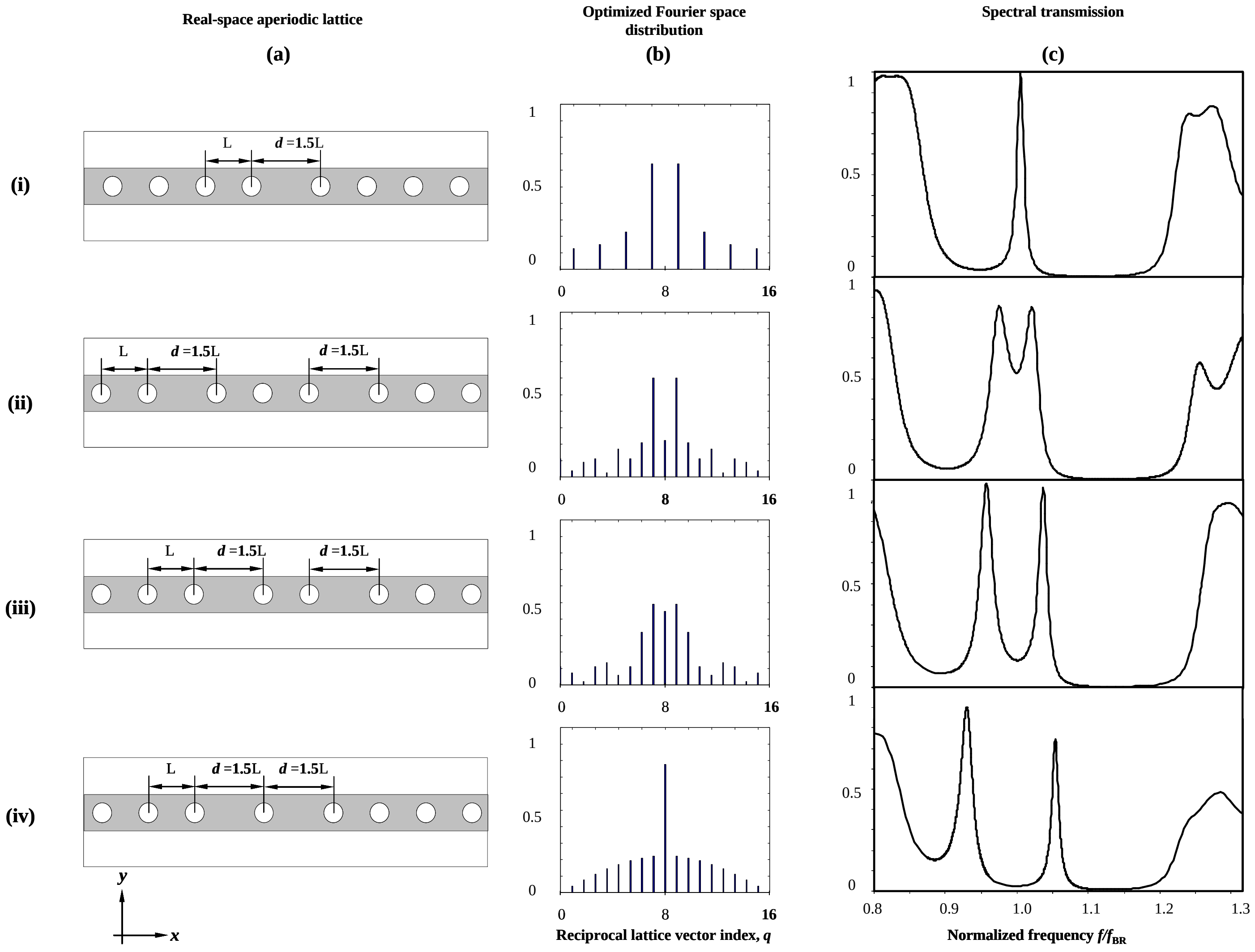

Fig.3.The Fourier component distributions and spectral transmission characteristics of four different aperiodic lattices doped with defects. We have used $N$=16 unit-cells, with the normalised DFT Nyquist frequency ($G_{BR}$) corresponding to $q$=$N$/2 = 8. Evident is the systematic progression from a single high-Q transmission peak to double high-Q transmission peaks.

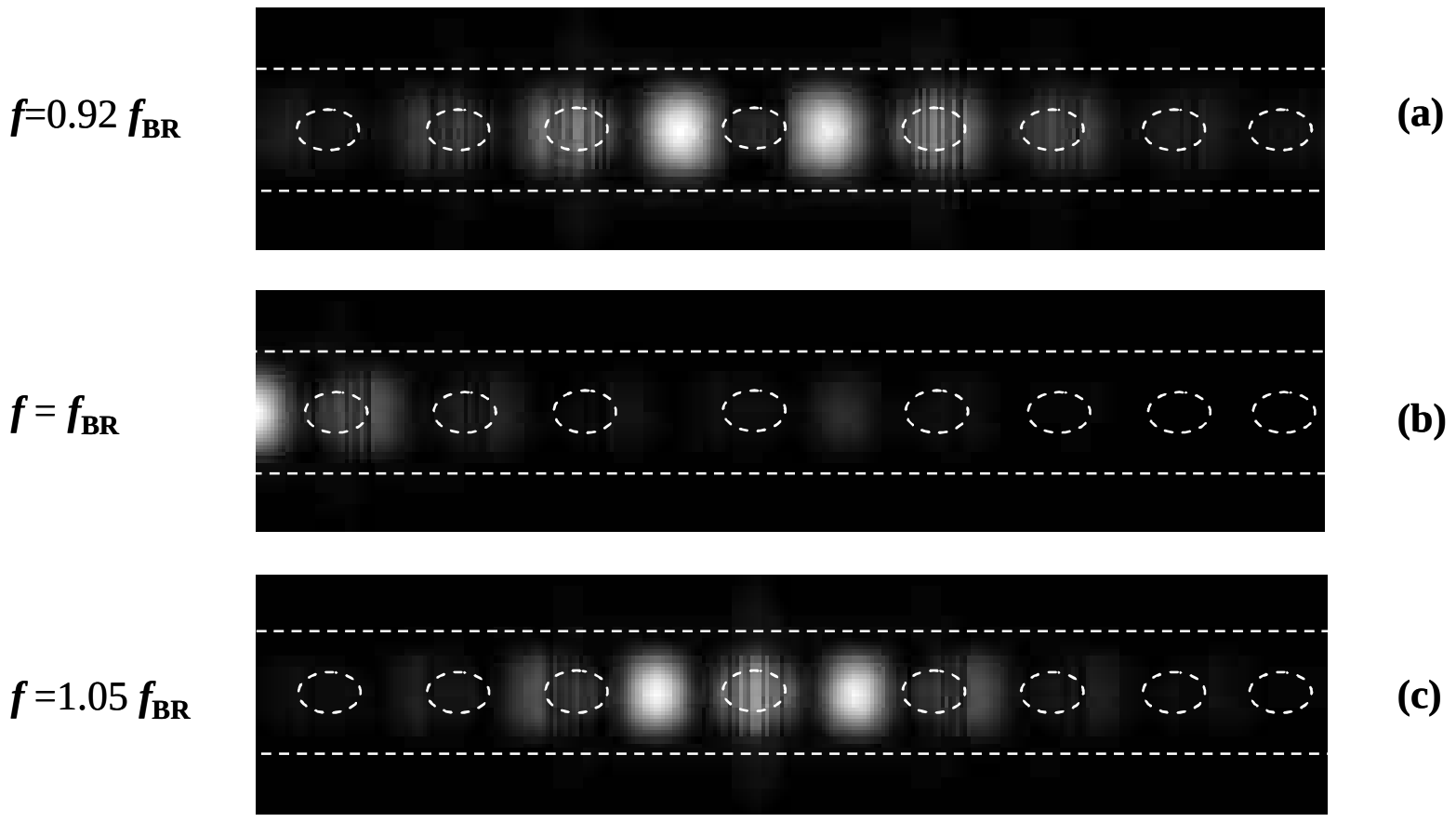

Fig.4. Simulated field intensity patterns in the *xy*-plane corresponding to the resonating modes (at frequencies $f$= 0.92 $f_{BR}$ and 1.05 $f_{BR}$) and at the Bragg frequency $f_{BR}$ for the aperiodic lattice of Fig.3 (iv-a).

FIG. 1: Qualitative understanding of how we can combine simple rules of DSP with Bragg's Law, in order to predict the position of the dips in the electromagnetic transmission spectrum of any real AL.

Figure 2. Optical transmission spectra (transmission $t$ versus normalized frequency $f/f_{BR}$), calculated using FIMMPROP-3D. Figures (i-v) show use of AL's to shift the PBG into the Brillouin Zone in steps of ~$0.02f_{BR}$ from the edge of the Bragg frequency $f_{BR}$.

Fig.3.The Fourier component distributions and spectral transmission characteristics of four different aperiodic lattices doped with defects. We have used $N$=16 unit-cells, with the normalised DFT Nyquist frequency ($G_{BR}$) corresponding to $q=N/2=8$. Evident is the systematic progression from a single high-Q transmission peak to double high-Q transmission peaks.

Fig.4. Simulated field intensity patterns in the *xy*-plane corresponding to the resonating modes (at frequencies $f$= 0.92 $f_{BR}$ and 1.05 $f_{BR}$) and at the Bragg frequency $f_{BR}$ for the aperiodic lattice of Fig.3 (iv-a).